\documentclass[conference]{IEEEtran}
\IEEEoverridecommandlockouts
\usepackage{cite}
\usepackage{amsmath,amssymb,amsfonts}
\usepackage{algorithmic}
\usepackage{graphicx}
\usepackage{textcomp}
\def\BibTeX{{\rm B\kern-.05em{\sc i\kern-.025em b}\kern-.08em
    T\kern-.1667em\lower.7ex\hbox{E}\kern-.125emX}}

\usepackage{graphicx}
\usepackage[T1]{fontenc}
\usepackage{lmodern}
\usepackage{textcomp}
\usepackage{latexsym}
\usepackage{bm}  
\usepackage{color}
\usepackage[normalem]{ulem} 
\usepackage{times,bm}
\usepackage{multicol}
\usepackage{graphicx}
\usepackage{epstopdf}
\usepackage{amsmath, amssymb}
\usepackage{type1cm}
\usepackage{amsmath,chemarrow}
\usepackage{cases}
\usepackage{pifont}

\newcommand{\pdif}[2]{\frac{\partial #1}{\partial #2}}

\newcommand{\cmark}{\ding{51}}%
\newcommand{\xmark}{\ding{55}}%
\newcommand{\RN}[1]{%
  \textup{\uppercase\expandafter{\romannumeral#1}}%
}

\begin{document}

\title{
Fault Diagnosis Method Based on Scaling Law for On-line Refrigerant Leak Detection
}

\author{\IEEEauthorblockN{Shun Takeuchi}
\IEEEauthorblockA{\textit{
Machine Discovery Technology Project
} \\
\textit{
Artificial Intelligence Laboratory
}\\
Fujitsu Laboratories Ltd., Kanagawa, Japan 
\\
takeuchi.shun@jp.fujitsu.com
}
\and
\IEEEauthorblockN{Takahiro Saito}
\IEEEauthorblockA{\textit{
Machine Learning Technology Project
} \\
\textit{
Artificial Intelligence Laboratory
}\\
Fujitsu Laboratories Ltd., Kanagawa, Japan 
\\
saitou.takah-02@jp.fujitsu.com
}
}

\maketitle

\begin{abstract}
Early fault detection using instrumented sensor data is one of the promising application areas of machine learning in industrial facilities.
However, it is difficult to improve the generalization performance of the trained fault-detection model because of the complex system configuration in the target diagnostic system and insufficient fault data. 
It is not trivial to apply the trained model to other systems.
Here we propose a fault diagnosis method for refrigerant leak detection considering the physical modeling and control mechanism of an air-conditioning system.
We derive a useful scaling law related to refrigerant leak.
If the control mechanism is the same, the model can be applied to other air-conditioning systems irrespective of the system configuration.
Small-scale off-line fault test data obtained in a laboratory are applied to estimate the scaling exponent.
We evaluate the proposed scaling law by using real-world data.
Based on a statistical hypothesis test of the interaction between two groups,
we show that the scaling exponents of different air-conditioning systems are equivalent.
In addition, we estimated the time series of the degree of leakage of real process data based on the scaling law and confirmed that the proposed method is promising for early leak detection through comparison with assessment by experts.
\end{abstract}

\begin{IEEEkeywords}
Fault detection and diagnosis, Soft sensor, Scaling law, Machine learning, Leak Detection
\end{IEEEkeywords}

\section{Introduction}
Failure of an industrial facility supporting social life causes not only the deterioration of operational efficiency, increase of labor cost for repair, and suspension of business but also affects the involved human life. 
The early detection of faults has attracted much attention as an important issue in the industry.
In particular, air-conditioning systems are equipment directly connected to our lives.
The typical fault type of the system is refrigerant leakage, which is caused by the cracking or deterioration of piping.
In addition hampering indoor temperature adjustment, 
refrigerant leak is well known to cause environmental contamination~\cite{koronaki2012refrigerant, mota2015commercial}.

Process monitoring is one of the effective methods for fault diagnosis~\cite{ge2012multivariate}.
Process variables, such as temperature, pressure, and flow rate, are measured by a large number of instrumented sensors 
(i.e., hardware or ``hard'' sensors) in industrial systems.
On the other hand, certain important process variables, such as the amount of refrigerant and product composition in distillation columns, are difficult or impossible to measure on-line owing to technical difficulties, time-consuming analyses, and/or the high cost of measuring devices. 
Soft sensor (or virtual sensor) techniques have been studied to monitor such difficult-to-measure process variables from easy-to-measure process variables by using machine learning techniques~\cite{fortuna2007soft}.
The soft sensor does not need to newly provide a measuring instrument dedicated to the target variable.

However, the essential problem in constructing a soft sensor for fault diagnosis is the insufficient training data~\cite{lou2002comparison}.
On the other hand, many facility systems such as air-conditioning systems have various system configurations.
Therefore, for the same target system, 
the statistical distributions between the training data and the test data are often not the same.
As a result, it is difficult to maintain a sufficient prediction accuracy of the target process.

To address this issue, 
we propose a fault diagnosis method using soft sensors by considering the physical model and control mechanism for detecting refrigerant leak in air-conditioning systems.
We present a scaling law that is useful for refrigerant leak detection, which is a power function that is satisfied between two process variables~\cite{mitzenmacher2004brief, newman2005power}. 
By considering the control mechanism of the target air-conditioning system in the basic equations for leakage, 
we derive the scaling law between the refrigerant mass and the refrigerant temperature, which is independent of the system configuration.
We thus build a soft sensor to estimate the degree of leakage from the refrigerant temperature by using the derived scaling law.

Our main contribution is the construction of a robust leak-detection model for air-conditioning systems with various  configurations.
In order to obtain the scaling law, we do not ``solve'' the basic equations for the target system.
It can be widely applied to industrial facilities such as air-conditioning systems.
Since the proposed soft sensor is constructed by the physics of the leakage~\cite{landau1987fluid}, 
it does not depend on the system configuration.
It is promising to apply the trained model for fault diagnosis in other diagnostic systems, which is the fundamental fault diagnosis approach in industrial facility systems.

The remainder of this paper is organized as follows.
In the next section, we discuss the related work.
We present the proposed leak detection method based on the scaling law in Section III.
The experiments and respective results are presented in Section IV.
The final section is devoted to concluding remarks.

\section{Related work}
\begin{figure}
\centerline{\
\includegraphics[width=88mm]{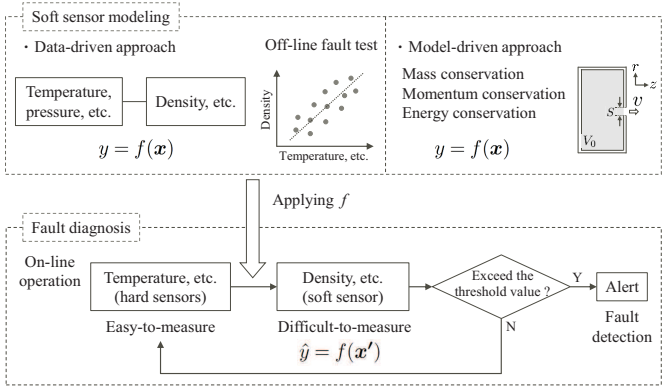}
}
\caption{
Schematic of the construction and operation of the soft sensor. 
}
\label{fig:softsensor}
\end{figure}

Soft sensor techniques are utilized to predict difficult-to-measure (or output) variable that is important for fault diagnosis from easy-to-measure (or input) variables (Figure \ref{fig:softsensor}).
There are two approaches to constructing soft sensors: data-driven soft sensors and model-driven soft sensors.


\subsection{Data-driven approach}
The data-driven soft sensor is constructed by using a machine learning technique~\cite{liu2010developing, serpas2013fault}.
The foundation of the data-driven soft sensor model is multiple linear regression analysis~\cite{bishop2006pattern}. The basic formula of the model is
\begin{eqnarray}
y_n^{(m)} = f^{(m)}(\bm{x}_n) = \bm{x}_n^{(m)} \bm{w}^{(m)} + \varepsilon^{(m)}, \label{eq:ddm}
\end{eqnarray}
where $y \in \mathbb{R}$ is the output variable, $\bm{x} \in \mathbb{R}^D$ are the input variables, $\bm{w}$ is the regression coefficient, $\varepsilon$ is the residual, $D$ is the dimensionality of the samples, and $n =1,2,\ldots,N^{(m)}$ is the number of observations. 
Facilities often have various operation modes.
Therefore, it is necessary to construct a soft sensor for each operation mode, ($m$).

The advantage of data-driven soft sensors is that a fault detection model can be constructed without expert knowledge of the target fault and system (e.g., the physics of refrigerant leakage)~\cite{beghi2016data, takeuchi2017semi}.

The drawback of the data-driven sensor model is the difficulty to maintain generalization performance with a lack of fault data for training.
Owing to quality improvement in recent years, equipment is hardly broken.
Therefore, in order to acquire training data, it is essential to reproduce the failure in a laboratory environment.
Meanwhile, since the configuration of large-scale complex facilities such as air-conditioning systems for business is designed according to the installation destination, the type of system configuration is not unique.
Therefore, it is inevitably necessary to build a model using fault data in a certain system configuration, apply the model to other systems, and diagnose the failure.
However, the correlation between input variables and the output variable depends on the system configuration.
In this situation, there is no guarantee that the correlation of diagnostic systems is the same, which makes the data-driven method unsuitable for application to diagnostic systems with other system configurations.

Here, we also refer to anomaly detection~\cite{chandola2009anomaly}.
While it does not require fault data in the training phase, there are multiple failure types in many industrial facilities, and the magnitudes of the failures are not unique.
In the case of a facility having many types of failures, 
since it is difficult to specify the type of failure only by the statistical deviance from the normal state,
an anomaly detection technique is not appropriate.

\subsection{Model-driven approach}
The model-driven soft sensor is constructed from the first principles based on physical properties~\cite{de1996model, kadlec2009data}.
Refrigerant leakage is a phenomenon in which refrigerant gas is ejected out of a container such as a pipe.
The properties of leakage are described by the hydrodynamic equations. 
We consider that fluids are non-dissipative and compressive; that is, the behavior of the fluid obeys Euler's equation. Considering that the leakage is caused by a tiny hole in the pipe, the system is described in the cylindrical coordinate $(r, \theta, z)$. We assume that the leakage occurs in the $z$ direction.
The basic equations are the continuity equation, the momentum equation, and the internal energy equation~\cite{landau1987fluid}:
\begin{align}
\pdif{\rho}{t} + \pdif{\rho v_z}{z} &= 0,  \label{eq:mass} \\
\pdif{\rho v_z}{t} + \pdif{(\rho v_z^2 + p)}{z} &= 0,  \label{eq:moment} \\ 
\pdif{e}{t} + \pdif{e v_z}{z} &= -p \pdif{v_z}{z} \label{eq:energy}.
\end{align}
Here, $\rho$ is the gas mass density, $\bm{v}$ is the leak velocity, $e$ [$=p/(\gamma-1)$] is the internal energy density of the gas, $p$ is the gas pressure, and $\gamma$ is the ratio of the specific heats, respectively. 
The set of equations (\ref{eq:mass}) -- (\ref{eq:energy}) is closed by the equation of state of a real gas: 
\begin{eqnarray}
p = z_c \rho R T,  \label{eq:eqstate}
\end{eqnarray}
where $z_c$ is the compressional constant and $R$ is the fluid constant.

By solving the basic equations (\ref{eq:mass}) -- (\ref{eq:eqstate}), it is possible to obtain the time evolution of the refrigerant gas.
As a promising model for a model-driven soft sensor of gas leaks, a method by using the pressure variation has been proposed~\cite{gu2006investigation, bergoglio2012leak, tian2016study}.
By integrating the continuity equation (\ref{eq:mass}) over the volume $dV$ ($= r^2 dr d\theta dz$), the loss rate of the gas mass density is expressed as
$ V \partial \rho/ \partial t = - \rho S v_z$.
Here, $S$ is the surface area of the hole, and $V$ is the volume of the hole.
For simplicity, we assumed that the gas density is constant in space.
By multiplying the above equation with the volume of the pipe in the target diagnostic system, $V_0$,
the time evolution of the refrigerant mass in the target diagnostic system, $M$ ($= \rho V_0$), is given by
\begin{eqnarray}
M(t) = M_0 \exp \left( - \frac{S}{V} \int_{t_0}^t v_z dt \right) \label{eq:mas},
\end{eqnarray}
where $M_0$ is the initial mass and $t_0$ is the starting time of leakage.
Assuming that the leakage velocity is constant in space, the compressional heating term $-p \partial v_z/\partial z$ is negligible in equation (\ref{eq:energy}).
For similarity with the derivation of equation (\ref{eq:mas}), 
the refrigerant pressure is given by
\begin{eqnarray}
p(t) = p_0 \exp \left( - \frac{S}{V} \int_{t_0}^t v_z dt \right). \label{eq:pre}
\end{eqnarray}
From equations (\ref{eq:mas}) and (\ref{eq:pre}), the relation between the refrigerant mass and pressure is obtained as
\begin{eqnarray}
\displaystyle  \frac{M(t)}{M_0} = \displaystyle  \frac{p(t)}{p_0}. \label{eq:convmodel}
\end{eqnarray}
The leakage is diagnosed from the variation of the gas pressure.

The advantage of model-driven soft sensors is that they do not depend on the system configuration, because they are obtained from the first-principles of the target fault.
The relation between input variables and the output variable is exactly determined without a training dataset.

However, it is difficult to diagnose refrigerant leakage in complex facility systems.
Various control mechanisms work to maintain the air-conditioning capability~\cite{legg2017air}.
These control mechanisms are designed under complex conditions and it is difficult to express them in differential equations.
Therefore, it is difficult to strictly solve the basic equation of the behavior of refrigerant leakage considering the actual operation environment, except in the case of a facility where operation is extremely simple.

\section{Proposed method}
In order to construct a soft sensor that maintains generalization performance even for facilities with various system configurations, 
our strategy is to build a model-driven soft sensor that takes into account the control mechanism of the facility.
It is not realistic to strictly solve the basic equation of the behavior in a complicated environment.
Therefore, we derive a scaling law that is useful for refrigerant leak detection.
The scaling law is a certain power function involving two process variables and is given by
$u_1 \propto u_2^k$,
where $k$ is the scaling exponent.
While equation (\ref{eq:convmodel}) is also the scaling law for $k=1$, it does not appropriately represent the behavior of real-world air-conditioning system.
We propose a scaling law with the unknown parameter of the control mechanism. 

\subsection{Diagnostic air-conditioning system and control mechanism}

\begin{figure}
\centerline{\
\includegraphics[width=69mm]{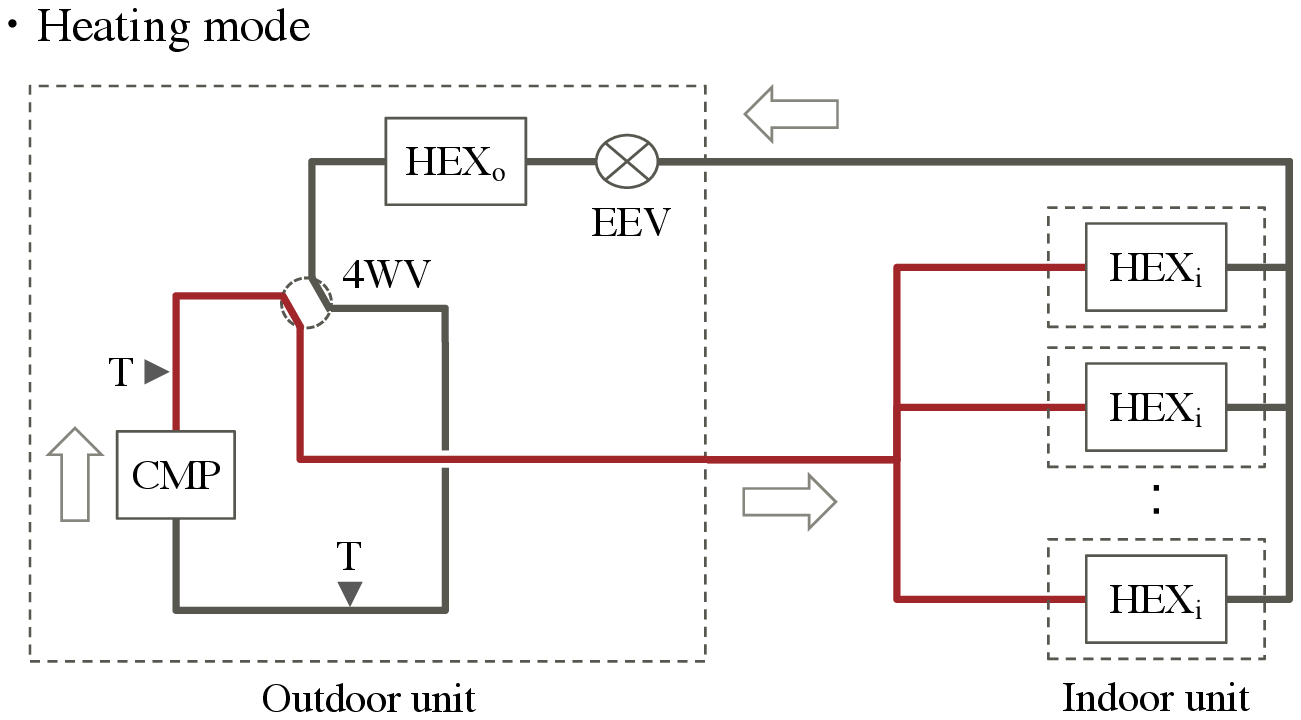}
}
\centerline{\
\includegraphics[width=69mm]{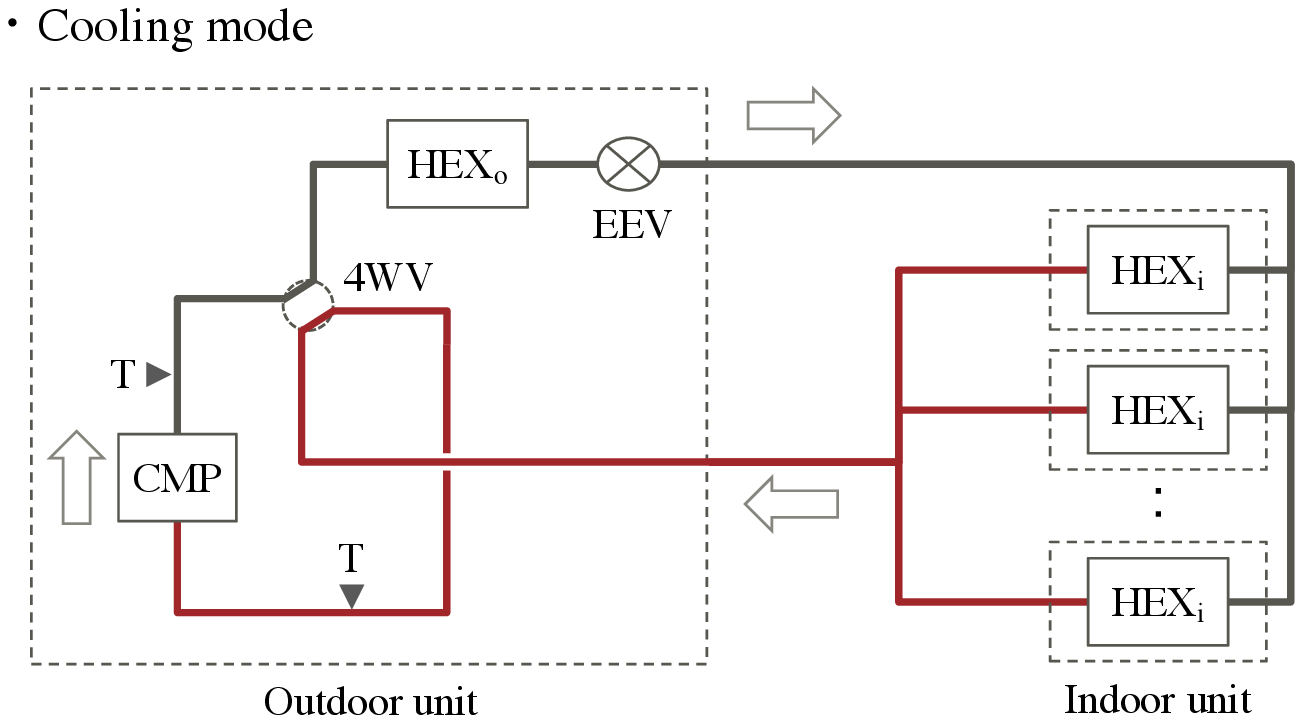}
}
\caption{
Configuration diagram of a diagnostic air-conditioning system.
The red piping indicates the discharge pipe (in the heating mode) and the intake pipe (in the cooling mode), in which a refrigerant temperature sensor is instrumented.
}
\label{fig:cycle}
\end{figure}

In this paper, we focus on the refrigerant leak in a commercial air conditioner known as packaged air conditioner as the  diagnostic target facility~\cite{legg2017air}.
We show the system configuration diagram in Figure \ref{fig:cycle},
where CMP is the compressor, HEX$_{\rm o}$ is the outdoor heat exchanger, HEX$_{\rm i}$ is the indoor heat exchanger, EEV is the electronic expansion valve, and 4WV is the four-way valve, respectively.
An outdoor unit consists of CMP, HEX$_{\rm o}$, EEV, and 4WV, and an indoor unit consists of HEX$_{\rm i}$.
These components are connected with refrigerant pipings. 
Commercial air conditioners have multiple indoor units according to the scale of the building.
Furthermore, in order to deal with facilities of various scales, they have various piping lengths, and the amount of refrigerant to be used differs for each system.

The air-conditioning system has two operation modes: heating operation and cooling operation.
The flow direction of the refrigerant is different in each mode.
The refrigerant receives heat transport and work in each component, and has different physical states in each piping.
Energy balance is satisfied in the refrigerating cycle.

In commercial air conditioners, various hard sensors are installed for each device in order to realize a stable air-conditioning environment.
The pipe on the outlet side of the CMP is called a discharge pipe, and the pipe on the inlet side of the CMP is called an intake pipe, which are shown in Figure \ref{fig:cycle} (red lines).
A temperature sensor for measuring the refrigerant temperature is installed in the pipes.

\begin{figure} 
\centerline{\
\includegraphics[width=70mm]{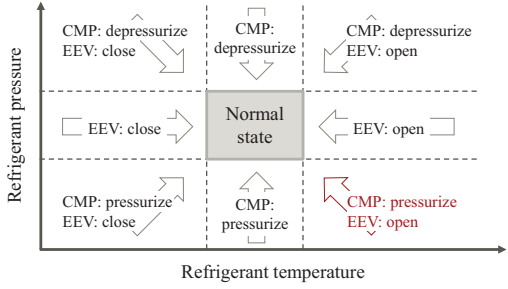}
}
\caption{
Control mechanism in an air-conditioning system related to refrigerant leak.
The reduction of refrigerant energy due to the leakage causes pressurization by CMP and refrigerant inflow by EEV opening (bottom right).
}
\label{fig:cntl}
\end{figure}

We describe the control mechanism necessary for the proposed method.
Figure \ref{fig:cntl} shows a schematic of a control mechanism in the air-conditioning system.
The air-conditioning system has a control mechanism of adjusting the refrigerant mass and the pressure in a specific piping so as to ensure necessary refrigerant energy to perform a desired heat exchange with the heat exchanger.
Since the heating mode and the cooling mode have different refrigerating cycles, the control mechanism also differs for each operation mode.
Note that this control mechanism is designed for normal operation, and the occurrence of this control does not directly indicate the occurrence of refrigerant leakage.

\subsubsection{Case of heating mode}
The CMP maintains the desired heat exchange.
The EEV adjusts the flow rate of the refrigerant and adjusts the distribution of the amount of refrigerant in the pipe in the air-conditioning system.
Pressurization/depressurization of the CMP and opening/closing of the EEV are performed so that the refrigerant pressure and the temperature of the discharge pipe attain a target value determined in the system design.
When refrigerant leakage occurs, the refrigerant energy decreases according to the energy conservation law.
Therefore, in order to compensate for the decrease in refrigerant energy due to leakage, pressure control is applied by the CMP, and the operation of refrigerant inflow by the EEV is expected to suppress the temperature rise due to pressure rise.

\subsubsection{Case of cooling mode}
Although the roles of CMP and EEV do not change even in the cooling mode, the piping to be controlled is an intake pipe, rather than a discharge pipe.

\subsection{Scaling law for refrigerant leak detection}
The basis of refrigerant leakage is equations (\ref{eq:mass}) -- (\ref{eq:eqstate}).
Here, we derive a basic equation that takes into account the control mechanism.
We focus on the physics in the discharge pipe and in the intake pipe.
In these pipes, in order to compensate for the decompression of the refrigerant due to the leakage, the refrigerant is replenished to the discharge pipe or intake pipe by opening the EEV, and pressurization is performed by the CMP.
Therefore, the mass conservation law (\ref{eq:mass}) and energy conservation law (\ref{eq:energy}) in these pipes are rewritten as
\begin{align}
\pdif{\rho}{t} + \pdif{\rho v_z}{z} &= C_1,  \label{eq:mass2} \\
\pdif{e}{t} + \pdif{e v_z}{z} &= -p \pdif{v_z}{z} + C_2. \label{eq:energy2}
\end{align}
Here, $C_1$ and $C_2$ are parameters resulting from the controls.
The discharge pipe and intake pipe are connected to the CMP and HEX$_{\rm i}$ respectively and the physical system of interest is an open system, but the effects of the inflow and outflow of the mass and energy are negligible since these terms are equivalent in the stable operation.

For similarity with the derivation of equation (\ref{eq:pre}), 
the term of compressional heating $-p \partial v_z/\partial z$ is negligible in equation (\ref{eq:energy2}).
By integrating over volume $dV$, the loss rates of the gas mass density and the pressure are
\begin{align}
\pdif{\rho}{t} &= - \rho \frac{S}{V} v_z + \frac{1}{V} \int C_1 dV, \label{eq:rhocont} \\
\pdif{p}{t} &= - p \frac{S}{V} v_z + \frac{\gamma -1}{V} \int C_2 dV. \label{eq:precont}
\end{align}
According to the nature of the control mechanism, each last term on the right-hand side of equations (\ref{eq:rhocont}) and (\ref{eq:precont}) works to compensate for the attenuation of the first term on the right-hand side.
Equations (\ref{eq:rhocont}) and (\ref{eq:precont}) are expressed as
\begin{align}
\pdif{\rho}{t} &= - (1-c_M) \rho \frac{S}{V} v_z, \label{eq:cont1} \\
\pdif{p}{t} &= - (1-c_p) p \frac{S}{V} v_z. \label{eq:cont2}
\end{align}
Here, $c_M$ ($0 \le c_M \le 1$) is the constant parameter of the control process for the EEV, and $c_p$ ($0 \le c_p \le 1$) is the constant parameter of the control process for the CMP.
The state without control ($C_1 = C_2 = c_M = c_p = 0$) is reduced to the conventional method described in Section II--B.
Since these controls are absolutely defined by the system design to maintain the normal state (Figure \ref{fig:cntl}), note that the parameters $c_M$ and $c_p$ themselves are unknown.

By using equation (\ref{eq:cont2}) and the equation of state (\ref{eq:eqstate}), we derive the equation for the refrigerant temperature:
\begin{eqnarray}
\pdif{T}{t} = (c_p-c_M) T \frac{S}{V} v_z. \label{eq:cont3}
\end{eqnarray}
By rearranging equations (\ref{eq:cont1}) and (\ref{eq:cont3}), 
we derive the following relation:
\begin{eqnarray}
\log \frac{M(t)}{M_0} = - \frac{1 - c_M}{c_p - c_M} \log \frac{T(t)}{T_0},
\end{eqnarray}
where $T_0$ is the initial temperature in the discharge pipe (for the heating mode) or in the intake pipe (for the cooling mode) and $M_0$ is the initial mass.
Since pressure control is performed in the target air-conditioning system,
we use the relationship between refrigerant mass and refrigerant temperature.
In this study, we do not aim to find the value of each coefficient for the control mechanism, $c_M$ and $c_p$.
Therefore, by defining $c = -(c_p - c_M)/(1 - c_M)$, we obtain the following relation:
\begin{eqnarray}
\frac{M(t)}{M_0} = \left[ \frac{T(t)}{T_0} \right]^{1/c}.  \label{eq:scale}
\end{eqnarray}
This is a scaling law for the target air-conditioning system with the control mechanism.

From this equation, it is possible to construct a soft sensor that estimates 
the amount of refrigerant $M(t)$ on-line.
In actual operation, there are air-conditioning systems in which the initial amount of refrigerant $M_0$ is unknown.
In many cases, it is sufficient to know the degree of leakage without knowledge of the amount of refrigerant.
We therefore diagnose refrigerant leak based on a soft sensor that measures the refrigerant leakage rate ($1-M(t)/M_0$); it is expressed by
\begin{eqnarray}
y = 1 - \frac{M(t)}{M_0} = 1 - \left[ \frac{T(t)}{T_0} \right]^{1/c}. \label{eq:model}
\end{eqnarray}

The proposed scaling law is independent of the physical properties and state of failure such as the compressional constant $z_c$ and the size or the configuration of the pipe crack.
It is derived from basic equations that hold for any system configuration.
Therefore, it is possible to transfer the predictive model to other systems with the same control process. 
If the control mechanism is the same, a change in the value of the scaling exponent $c$ is not expected.
The scaling exponent is estimated through machine learning with a training fault dataset. 
Note that the parameter $c$ has physical rationale.

The advantage of the proposed scaling law for fault diagnosis is the scale invariance of the process variables~\cite{mitzenmacher2004brief, newman2005power}.
Equation (\ref{eq:scale}) has the same power function $(T/T_0)^{1/c}$ from a small scale air-conditioning system with a small initial refrigerant mass to a large scale air-conditioning system with a large initial refrigerant mass.
In other words, in order to estimate the parameter $c$, we can use the data of a small scale air-conditioning system in a laboratory.
This is convenient in industrial facility systems with various system scales.

\subsection{Procedure of proposed method}
We describe the procedure of the proposed method, which has four phases.

\subsubsection{Estimating the scaling exponent}
The scaling exponent is estimated in this phase.
The air-conditioning system used in this phase may be a system with a configuration 
different from that of the system to be diagnosed if the control mechanism are the same.
The refrigerant mass $M$ should be measurable.
In this study, we consider a laboratory system.
In the laboratory environment, the initial refrigerant mass of the system $M_0$ is known.
By measuring the leaked mass, it is possible to estimate the refrigerant mass $M$ at the time.
The temperature $T$ can be obtained from the instrumented hard sensor.
In this case, the explanatory variable is the refrigerant mass to be leaked,
and the objective variable is the refrigerant temperature.
Therefore, by transposing the refrigerant mass and the temperature of the equation (\ref{eq:scale}),
we use the relation
\begin{eqnarray}
\log \frac{T(t)}{T_0} = c \log \frac{M(t)}{M_0}, \label{eq:scale3}
\end{eqnarray}
for estimating the scaling exponent $c$.
The parameter $c$ is trained using linear regression analysis~\cite{bishop2006pattern}.
This procedure is performed separately for the heating mode and the cooling mode.

\subsubsection{Calculating the initial temperature in the diagnostic system}
In this phase, by using the dataset of the target diagnostic system, the initial temperature $T_0$ is calculated before starting the diagnosis.
In an actual operation environment, the refrigerant temperature has dispersion due to complicated heat exchange and controls.
A certain data-collection period is provided, and the initial temperature is calculated.
This temperature is the refrigerant temperature corresponding to the current operation mode, that is, discharge temperature (for the heating mode) or intake temperature (for the cooling mode).

\subsubsection{Estimating the degree of refrigerant leak in the diagnostic system}
In this phase, leakage diagnosis is performed using the scaling exponent $c$ and initial temperature $T_0$ obtained in Phases 1 and 2.
The dataset of the system used in Phase 2 is used in this phase.
From the acquired refrigerant temperature $T$, the degree of refrigerant leakage $y$ is estimated on-line using equation (\ref{eq:model}).
Here, the initial degree of refrigerant leakage $y_0$ is calculated using the amount of refrigerant charged in the system, if it is known.
If it is unknown, it is assumed that the system is in the normal state $y_0=0$ at the start of diagnosis.
The temperature is the refrigerant temperature corresponding to the current operation mode.
Based on the estimated degree of refrigerant leakage $\hat{y}$, refrigerant leakage is diagnosed.

\subsubsection{Updating parameters at the operation-mode change}
In real-world operation, the operation mode of the air-conditioning system might be switched.
In this case, it is necessary to update the initial degree of refrigerant leakage $y_0$, the initial temperature $T_0$, and the scaling exponent $c$ in the operation mode after switching.
Refrigerant leakage is independent of the switching of operation modes.
The degree of refrigerant leak at the time when the operation mode is switched is the degree of  refrigerant leakage just before switching.
That is, the following relation holds at the time of switching of the new operation mode:
\begin{eqnarray}
1 - \frac{M_{0}^{\RN{2}}}{M_0} = y^{\RN{1}}, 
\label{eq:modechang} 
\end{eqnarray}
where superscript $\RN{2}$ corresponds to the new operation mode and 
$\RN{1}$ corresponds to the previous operation mode.
Since the scaling law of equation (\ref{eq:scale}) is satisfied for any amount of refrigerant, the degree of leakage  is estimated by using the scaling law of the initial mass in the new operation mode $M_ {0}^{\RN{2}}$.
Thus, by using equations (\ref{eq:model}) and (\ref{eq:modechang}), the relation of the degree of leakage considering the switching of new operation modes is given by
\begin{eqnarray}
y^{\RN{2}} = 1 - \frac{M}{M_0} = 1 - \left[ 1 - y^{\RN{1}}  \right] \left[ \frac{ T(t) }{ T_0 } \right]^{ 1/c }, \label{eq:model_change}
\end{eqnarray}
where, the refrigerant temperature $T(t)$, the initial temperature $T_0$, and the scaling exponent $c$ are the values in the new operation mode.
The initial temperature in the operation mode after switching is calculated based on the procedure of Phase 2, 
and the scaling exponent has the value trained in Phase 1.
When the operation-mode returns to the former mode, the initial mass and temperature in that mode are already calculated.
In this case, the degree of leakage is estimated by using equation (\ref{eq:model}).

\section{Experiments}
In this section, we present the experimental results of applying the proposed soft sensor model to real-world operation data.

\begin{table*}
 \begin{center}
 \caption{Real-world datasets of air-conditioning systems.}
 \label{table:expdata}
  \begin{tabular}{ccccccc}
   \hline
    Dataset & Description & Operation mode &No. of indoor units & Initial refrigerant mass & Temperature & Refrigerant mass \\
   \hline
    A & Fault test data & Heating, cooling & 3 & 18 kg & \cmark & \cmark \\
    B & Fault test data & Heating, cooling & 14 & 48 kg & \cmark & \cmark \\
    C & Real process data & Heating, cooling & 7 & Unknown & \cmark & \xmark \\
   \hline
  \end{tabular}
\end{center}
\end{table*}

\subsection{Real-world datasets of air-conditioning system}
We used three datasets to validate our method (Table \ref{table:expdata}).
These datasets are from three air-conditioning systems with different configurations.
The values of process variables such as refrigerant amount and refrigerant temperature are different among these systems.
Note that the control mechanism of the air conditioner is the same.

Dataset A is an off-line fault test obtained in a laboratory.
The refrigerant leakage of the system is reproduced by artificially withdrawing the refrigerant from the piping during operation.
The refrigerant mass of the system can be measured from the initial mass and leaked mass\footnote{
In all the air-conditioning systems (datasets A, B, and C) used in this experiment, the intake pipe branches into two, and the refrigerant temperature of each piping is measured.
Therefore, as the intake temperature in this experiment, the average value of the refrigerant temperatures of the two pipes is adopted.
}.
This fault test is carried out in each of the heating mode and the cooling mode in which the operation is different.

Dataset B is similar to dataset A, but its system has a different configuration.
Compared with the system of dataset A, the system of dataset B is larger.
In order to reproduce more realistic refrigerant leakage, the fault test was conducted in an environment in which the refrigerant gradually leaks because of the refrigerant pressure of the system.

Dataset C is from an air-conditioning system in an actual operation environment.
The system configuration and installation environment are different from those in the laboratory.
Since it is not possible to obtain the system installation information, the initial refrigerant mass $M_0$ is unknown.
Dataset C contains operation-data logs from December 2014 to March 2016, and the system operates in the heating mode in winter and in the cooling mode in summer.
No hard sensor for measuring the refrigerant mass is mounted.
It is known from the experiential assessment of experts that refrigerant leakage occurred in the air-conditioning system of dataset C.
The system was stopped for repair on April 2016.

\subsection{Evaluating the proposed scaling law}
We showed the useful scaling law related to leakage.
This relation is derived under the assumption that the behavior of the control mechanism is constant: $c_M = c_p = {\rm constant}$.
Here, by using the fault test datasets A and B, we verify that the proposed scaling rule is satisfied.

\begin{figure} 
\centerline{\includegraphics[width=80mm]{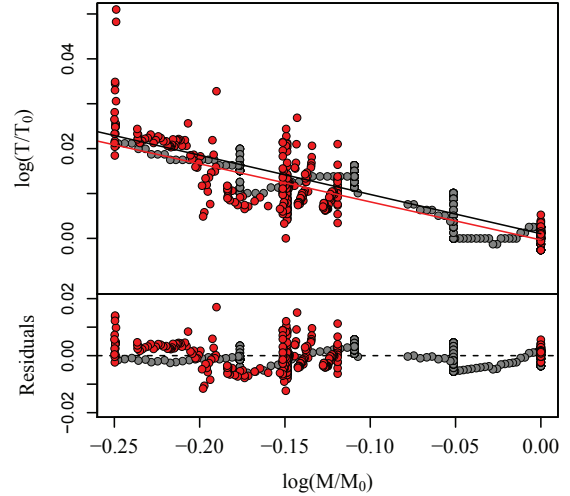}}
\caption{
Scatter plot of refrigerant mass and refrigerant temperature (upper panel) and that of residuals (lower panel) drawn using datasets A and B in the heating mode.
The black dots indicate the dataset A, and the red dots indicate the dataset B.
The straight line of the upper panel shows the regression line trained with each dataset.
}
\label{fig:scalecheck_h}
\end{figure}

Figure \ref{fig:scalecheck_h} shows scatter plots drawn using datasets A and B in the heating mode.
The upper panel is a scatter plot of the refrigerant mass and the temperature corresponding to equation (\ref{eq:scale3}).
The refrigerant temperatures indicate the discharge temperature.
The black dots correspond to dataset A, and the red dots correspond to dataset B.
The regression lines in each dataset are shown in the upper panel.
In order to evaluate the linearity of both regression lines, a scatter plot of residuals against the refrigerant mass ratio is shown in the lower panel.
The residual is roughly constant with respect to each refrigerant mass ratio. 
That is, both datasets have constant slopes.
In the range where the residual is not constant (e.g., $-0.17 \lesssim \log(M/M_0) \lesssim -0.12$ in dataset A), it shows the transient state during the rapid refrigerant leakage.
Before the air-conditioning system attains the normal state (Figure \ref{fig:cntl}), the residuals do not become constant, because of leakage.
In the range where the system is stable (e.g., $\log(M/M_0) \sim -0.1$ or $-0.17$ in dataset A), the residuals are stable.

The figure shows that, although the two air-conditioning systems have different configurations, they have similar slopes.
Table \ref{table:scalecoef} lists the slope (that is, scaling exponent) and intercept of the regression line trained by each dataset.
Although the proposed scaling law does not include the intercept, it is taken into account as a measurement error.
The important point is that the scaling exponent does not depend on the system configuration.
To verify that the scaling exponents obtained for each data set are equivalent, a test of homogeneity of slopes is performed.
The test is used for the analysis of covariance and evaluates the statistical significance of interaction between groups.
The null hypothesis is that the slopes of the two straight lines are equal, and it is a passive judgment (not showing parallelism); however, it is used as a test of parallel lines.
We adopt Student's t test for verifying between the two groups in this verification~\cite{wilkinson1973symbolic}.
The result of the parallel-line analysis is summarized in Table \ref{table:parallel}.
In the heating mode, the statistics determined for inspection does not exceed a level of significance of $5\%$, and the parallelism of the inclination of the two groups cannot be denied.
Therefore, in the heating mode, it is claimed that the scaling law of equation (\ref{eq:scale}) does not depend on the air-conditioning system.

\begin{figure} 
\centerline{\includegraphics[width=80mm]{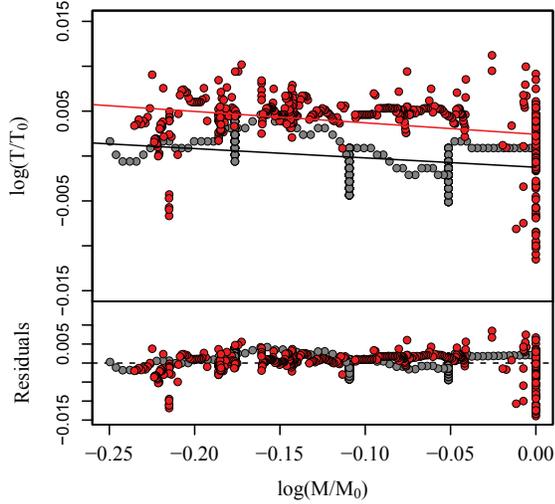}}
\caption{
Same as Fig. \ref{fig:scalecheck_h}, but in cooling mode.
}
\label{fig:scalecheck_c}
\end{figure}

Figure \ref{fig:scalecheck_c} shows the results in the cooling mode.
As in the heating mode, it is found that each dataset follows the inclination of certain slopes.
Table \ref{table:scalecoef} and \ref{table:parallel} show the coefficient (scaling exponent) and intercept of the regression line in the cooling mode, as well as the results of parallel-line analysis.
From Table \ref{table:parallel}, even in the cooling mode, it is claimed that equation (\ref{eq:scale}) does not depend on the system configuration.

\begin{table}
 \begin{center}
 \caption{Scaling exponent in each dataset.}
 \label{table:scalecoef}
  \begin{tabular}{cccc}
   \hline
    Operation mode & Dataset & Scaling exponent, $c$ & Intercept \\
   \hline
    Heating & A & -0.0874 &  0.00105 \\
    & B & -0.0826 & -0.000170 \\
   \hline
    Cooling & A & -0.0104 & -0.00125 \\
    & B & -0.0127 &  0.00242 \\
  \hline
  \end{tabular}
\end{center}
\end{table}

\begin{table}
 \begin{center}
 \caption{Statistics for test of the interaction between two group.}
 \label{table:parallel}
  \begin{tabular}{ccc}
   \hline
    Operation mode & $t$ value & $p$ value \\
   \hline
    Heating & 0.696 & 0.486 \\
    Cooling & -1.164 & 0.244 \\
  \hline
  \end{tabular}
\end{center}
\end{table}

In each operation mode, the scaling exponent is $c < 0$.
When the refrigerant leaks, the refrigerant energy is reduced according to the energy conservation law.
Since the refrigerant temperature is compressed to the target value by the pressure control of the CMP, the refrigerant temperature of the discharge pipe or intake pipe is increased.
On the other hand, since the refrigerant is simultaneously supplied to the discharge pipe or intake pipe by the EEV, the rise in the refrigerant temperature is suppressed.
As shown by equation (\ref{eq:cont3}), when there is no control ($c_p = c_M = 0$), the refrigerant temperature remains constant regardless of refrigerant leakage.
The proposed scaling law is appropriately formulated by considering the leakage of a real-word air-conditioning system.

When we omit the intercept, the parallelism is denied in both operation modes.
Since the air-conditioning system operates under complicated control mechanism, 
the refrigerant temperature ratio contains the components of fluctuation.
Suppression of the fluctuation  is a future work.

Here, we supplement the refrigerant leakage in idle mode of the air-conditioning system.
The real-world data has an idle mode in addition to the heating mode and the cooling mode.
In the idle mode, the control mechanism described in Section III does not work.
Refrigerant leakage can be detected in principle by the conventional method, which does not consider the control mechanism in this state.
However, since the refrigerant does not circulate in the system in this state, the influence of the refrigerant leak on each process variable is restricted to the vicinity of cracking or deterioration of the piping.
Therefore, if there is no pressure sensor in the vicinity of the site of the refrigerant leakage, the refrigerant leak cannot be detected.
The hard sensor is installed in a limited place in the system, and it is not realistic to detect the refrigerant leakage from the pressure loss in the idle state.

\subsection{Estimating the degree of leakage}
\begin{figure} 
\centerline{\
\includegraphics[width=90mm]{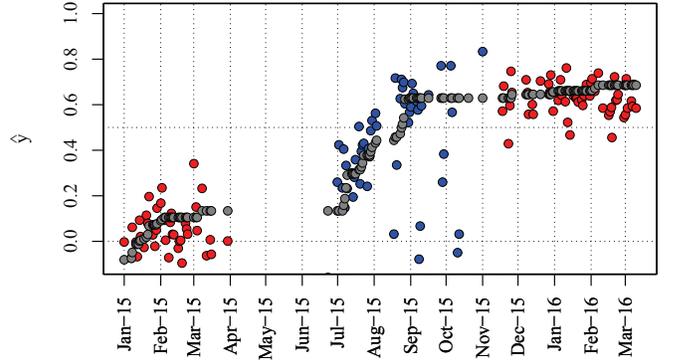}
}
\caption{
Time series of the degree of refrigerant leak estimated for dataset C. 
The red dots correspond to the heating mode, and the blue dots correspond to the cooling mode.
The black dots indicate the degree of leakage considering moving average and monotonic increase.
Through expert assessment, it was detected in April 2016 that refrigerant leakage occurred in the air-conditioning system.
}
\label{fig:timeseries}
\end{figure}

We found that the proposed scaling law was satisfied with real-world data.
Here, refrigerant-leakage estimation is verified by using dataset C, which contains real process data with refrigerant leakage assessed by an expert.
The operation mode is switched from the heating mode to the cooling mode and further to the heating mode.
Therefore, equation (\ref{eq:model}) is applied to the heating mode and equation (\ref{eq:model_change}) is applied to the cooling mode.
For the training of the scaling exponent $c$ in each operation mode, dataset A is used
in this experiment.
Even if dataset B is used, the result does not change.
The data collection period for estimating initial temperature is adopted 1 week.
The refrigerant temperature for estimating the leakage is adopted as its average value for each day.

Figure \ref{fig:timeseries} shows the time series of the soft sensor of the degree of leakage.
Here, we assumed that the system was in the state with no leakage ($y_0=0$) at the start of data collection.
The red and blue dots correspond to the refrigerant leakage in the heating and cooling mode, respectively.
The degree of refrigerant leak can be obtained from the refrigerant temperature in each operation mode.
We confirmed that, in spite of the independent estimation, the degrees of leakage in the cooling mode and in the heating mode are equivalent at the operation mode change in November 2015.
That indicates that the proposed scaling law and the trained scaling exponents in each operation mode are appropriate.

The refrigerant temperature has a large dispersion due to complicated control and heat exchange.
The variance of the estimate interferes with the appropriate diagnosis.
Here, we consider the 1-week moving average and the monotonicity of the leak~\cite{han2011data}.
Since the refrigerant pressure is greater than the ambient pressure, the refrigerant does not return to the pipe.
The result of considering the moving average and the monotonous increase in refrigerant leakage with respect to time is the black dots of Figure \ref{fig:timeseries}.
It is clear that the leakage of refrigerant is increasing with time.
Refrigerant leakage was detected by experts in April 2016 in this system.
If the refrigerant leak does not occur sufficiently, it is difficult for a human to detect the anomalous state of air-conditioning performance because of the control mechanism.
For example, when $\hat{y} = 0.5$ is set as the threshold value, the leakage is detected in September 2015, which is sufficiently earlier than the detection by a human.
In this air-conditioning system, it is inferred that refrigerant leakage occurred during the cooling mode.
On the other hand, the estimation of leakage considering monotonic increase is influenced by the outlier value of refrigerant temperature. 
Furthermore, 
environment temperature variations would influence the refrigerant temperature, bringing noise to the measured data.
A future work is to devise a detection rule and the threshold value assuming actual operation.

\section{Conclusions}
In this paper, we have proposed a fault diagnosis method aimed at detecting refrigerant leak in commercial air-conditioning systems.
We have derived the scaling law related to refrigerant leakage, taking into consideration the control mechanism of the target air-conditioning system in the basic equations expressing refrigerant leak.
Since the proposed scaling law is derived from the physics of the leakage, it is a relation that does not depend on the system configuration.
With the proposed method, it is possible to realize fault diagnosis by constructing a soft sensor without encountering the problem of insufficient fault data for training which occurs in machine fault diagnosis.
Our data analysis method, which designs a soft sensor model based on the target domain knowledge and estimates parameters that are complicated and difficult to formulate by machine learning techniques, is a realistic approach for dealing with real-world constraints.

\section*{Acknowledgement}
We would like to thank the Office of Air Conditioner Products Development of Fujitsu General Limited for providing us with the air-conditioning system datasets.

\bibliographystyle{ieeetran}
\bibliography{ref}

\end{document}